\documentclass[jkps,preprint,fleqn,showpacs,showkeys]{revtex4}
\usepackage{graphicx}
\usepackage{subfigure}
\usepackage{amssymb}
\usepackage{amsmath}
\usepackage{bm}
\begin{document}
\setcounter{page}{0}
\title[]{Critical properties of a spin-1 triangular lattice Ising antiferromagnet}
\author{Milan \surname{\v{Z}ukovi\v{c}}}
\email{milan.zukovic@upjs.sk}
\author{Andrej \surname{Bob\'ak}}
\affiliation{Department of Theoretical Physics and Astrophysics, Faculty of Science,\\ 
P. J. \v{S}af\'arik University, Park Angelinum 9, 041 54 Ko\v{s}ice, Slovak Republic}

\date[]{Received 25 May 2012}

\begin{abstract}
We employ Monte Carlo simulations in order to investigate critical behavior of a geometrically frustrated spin-1 Ising antiferromagnet on a triangular lattice in the presence of a single-ion anisotropy. It has been previously found that long-range order can exist in the isotropic system with a spin larger than some critical value estimated as 11/2. We show that the presence of the single-ion anisotropy can lead to a partial long-range order in the low-temperature region even below this critical value, namely for the spin 1, within a certain range of the anisotropy strength. At higher temperatures we identify another phase of the Berezinsky-Kosterlitz-Thouless type and using a finite-size scaling analysis evaluate the correlation decay exponent. We also study densities of various local spin patterns in the respective phases.
\end{abstract}



\maketitle

\section{INTRODUCTION}

The triangular lattice Ising antiferromagnet (TLIA) with the spin 1/2 is well known to display no long-range ordering down to zero temperature~\cite{wann} where the spin-correlation function decays as power-law~\cite{step}. On the other hand, in the ground state a long-range order with the partially ordered structure, characterized by two sublattices of opposite magnetizations and one sublattice of zero magnetization, can occur if the spin is larger than some critical value~\cite{naga,yama,lipo}. The upper bound of this critical value was estimated by the use of Peierls' argument~\cite{naga} as 62 and the precise value was established by Monte Carlo simulations~\cite{yama} as 11/2. The emergence of the long-range order was argued to arise as a result of a drastic change in the ground-state degeneracy as well as some other physical quantities of this frustrated spin system with the increasing spin magnitude. Nevertheless, the large degeneracy can be lifted by some perturbations, such as an external magnetic field~\cite{metcalf,schick,netz} or selective dilution~\cite{kaya,zuko}, which have been shown to lead to long-range ordering even in the highly frustrated spin-1/2 system. For Ising models with spin larger that 1/2 a single-ion anisotropy is another parameter that may play a crucial role in their critical properties (see, e.g.,~\cite{cape,blum,beal}). 

In the present paper we show that this is also the case for the current model and demonstrate how the inclusion of the single-ion anisotropy can change the above scenario. We show that even a small amount of the single-ion anisotropy leads to a partially ordered phase at low temperatures, while different critical phase of the Berezinsky-Kosterlitz-Thouless (BKT) type~\cite{kost}, characterized by a power-law decay of the spin correlation function, persists at higher temperatures.

\section{SIMULATION DETAILS}
We consider the Ising model described by the Hamiltonian 
\begin{equation}
\label{Hamiltonian}
H=-J\sum_{\langle i,j \rangle}S_{i}S_{j}-D\sum_{i}S_{i}^2,
\end{equation}
where $S_{i}=\pm1,0$ is an Ising spin on the $i$th lattice site, $\langle i,j \rangle$ denotes the sum over nearest neighbors, $J<0$ is the antiferromagnetic exchange interaction parameter, and $D$ is the single-ion anisotropy parameter. 

We perform standard Monte Carlo (MC) simulations following the Metropolis dynamics. We simulate the triangular lattice spin systems of the linear size $L$ and apply the periodic boundary conditions. For thermal averaging we typically consider $N=2 \times 10^5$ MCS (Monte Carlo sweeps or steps per spin) after discarding another $N_{0} = 0.2 \times N$ MCS for thermalization. 
In order to obtain dependencies of various thermodynamic quantities on the reduced temperature $t=k_BT/|J|$, we use $L=48$ and start simulations from high temperatures with random initialization. Then the temperature is gradually lowered with the step $\Delta t=0.02$  and the simulations start from the final configuration obtained at the previous temperature. In order to estimate the error bars, we run five simulations with different initializations. For finite-size scaling analysis (FSS), employed to identify different phases, we use the lattice sizes $L=24,48,72,96$ and $120$.

We calculate the following quantities. The sublattice magnetization of the sublattice X (X=A,B and C) is obtained as $M_{\mathrm{X}}=\sum_{i \in X}S_{i}$. Then, for an antiferromagnet, as an order parameter it is useful to define the staggered magnetization per site as
\begin{equation}
\label{mag_st}
m_s = \langle M_s \rangle/L^2 = 3\langle \max(M_{\mathrm{A}}, M_{\mathrm{B}}, M_{\mathrm{C}})-\min(M_{\mathrm{A}}, M_{\mathrm{B}}, M_{\mathrm{C}}) \rangle/2L^2,
\end{equation}
\noindent where $\langle\cdots\rangle$ denotes thermal average. Further, from fluctuations of the above quantities we define the specific heat per site $c$ as
\begin{equation}
\label{eq.c}c=\frac{\langle H^{2} \rangle - \langle H \rangle^{2}}{L^2k_{B}T^{2}},
\end{equation}
and the staggered susceptibility per site $\chi_{s}$ as
\begin{equation}
\label{eq.chi}\chi_{s} = \frac{\langle M_{s}^{2} \rangle - \langle M_{s} \rangle^{2}}{L^2k_{B}T}.
\end{equation}

In the ground state the spin-correlation function of the TLIA model decays as a power law~\cite{step}
\begin{equation}
\label{PL}\langle S_{i}S_{j} \rangle \propto r_{ij}^{-\eta},
\end{equation}
where the exponent $\eta$ has been shown to decrease with the spin value from $\eta=1/2$ for spin-1/2 to zero for spin larger than 11/2, for which antiferromagnetic long-range order occurs~\cite{naga,yama}. The value of $\eta$ can be estimated by FSS of the order parameter $m_s$, which scales as~\cite{chal}
\begin{equation}
\label{ms_FSS}
m_s(L) \propto L^{-\eta/2}
\end{equation}
or the quantity~\cite{naga,yama}
\begin{equation}
\label{Y}
Y = \Big\langle\Big[\Big(\sum_{i \in A}S_{i}\Big)^2+\Big(\sum_{j \in B}S_{j}\Big)^2+\Big(\sum_{k \in C}S_{k}\Big)^2\Big]\Big\rangle/L^2,
\end{equation}
which scales as
\begin{equation}
\label{Y_FSS}
Y(L) \propto L^{2-\eta}.
\end{equation}

\section{RESULTS AND DISCUSSION}
Let us first examine the ground-state properties for different values of $D$. For $D=0$, the energetic arguments dictate that the ground-state configuration is such that the spins on each elementary triangular plaquette sum to $\pm 1$. All such plaquettes are energetically equivalent with the energy per spin $e=J$ and there is no long-range order among them. It is easy to verify that for $D>0$ the preferred configurations are such that the spins on each elementary triangular plaquette again sum to $\pm 1$ but zero spin states are not involved. Then the energy per spin is $e=J-D$ and the system behaves like a spin-1/2 Ising model with no long-range order~\cite{wann}. If $D<0$ the configurations with the spin $S_{i}=\pm1$ states on one sublattice are suppressed and the partially ordered phase with the antiferromagnetic ordering on two sublattices and non-magnetic states of the third sublattice, i.e., the $S_{i}=0$ states, can occur. Such configurations are characterized by the energy per spin $e=J-2D/3$. However, for $D<3J/2$ the energy becomes positive and therefore the non-magnetic state with all spins taking zero value is the ground state. Therefore, the partial long-range order can only be expected within $3J/2<D<0$.

\begin{figure}[t]
\centering
		\subfigure{\includegraphics[scale=0.42]{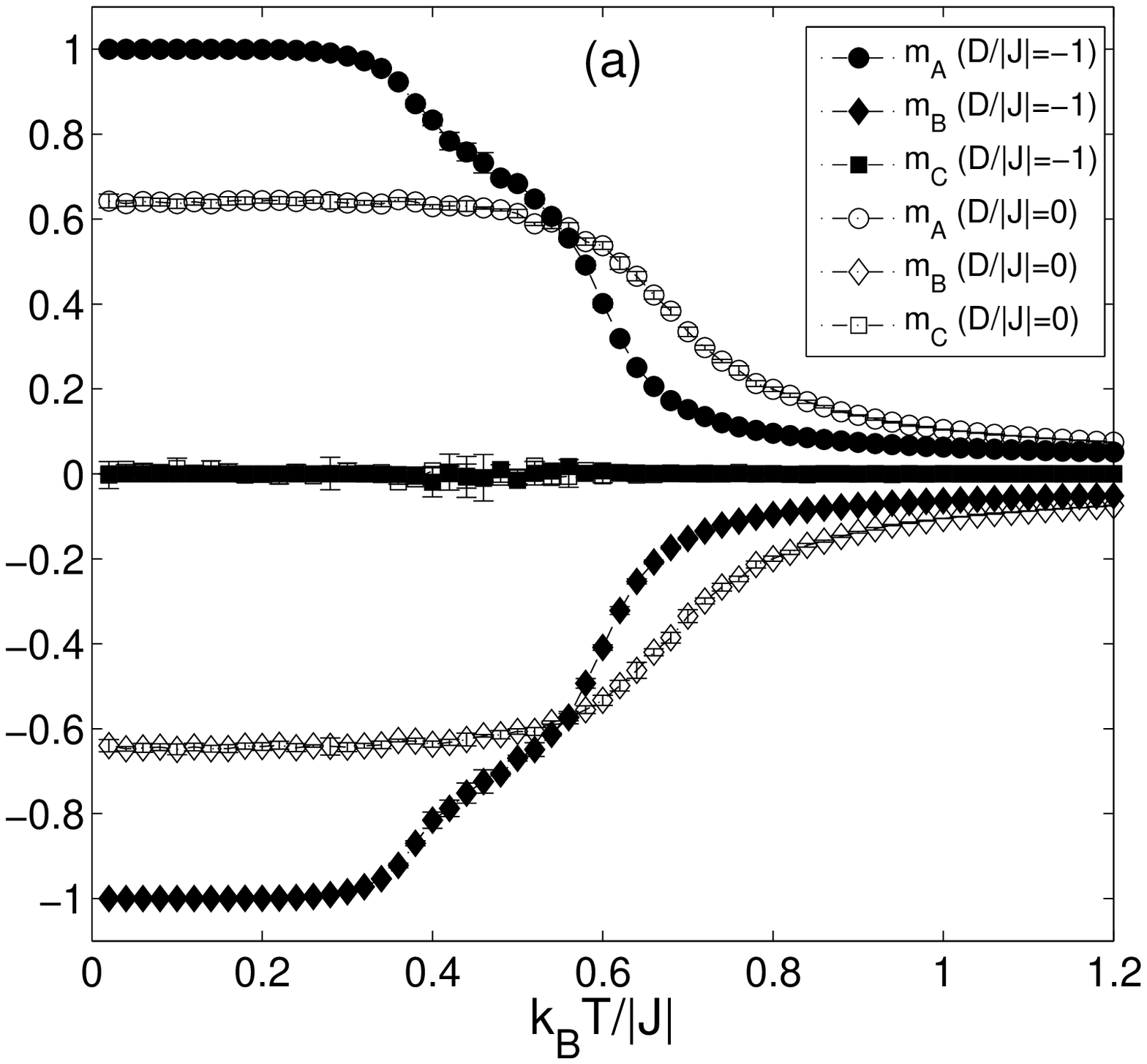}\label{fig:mi-T_L48}}\\
    \subfigure{\includegraphics[scale=0.42]{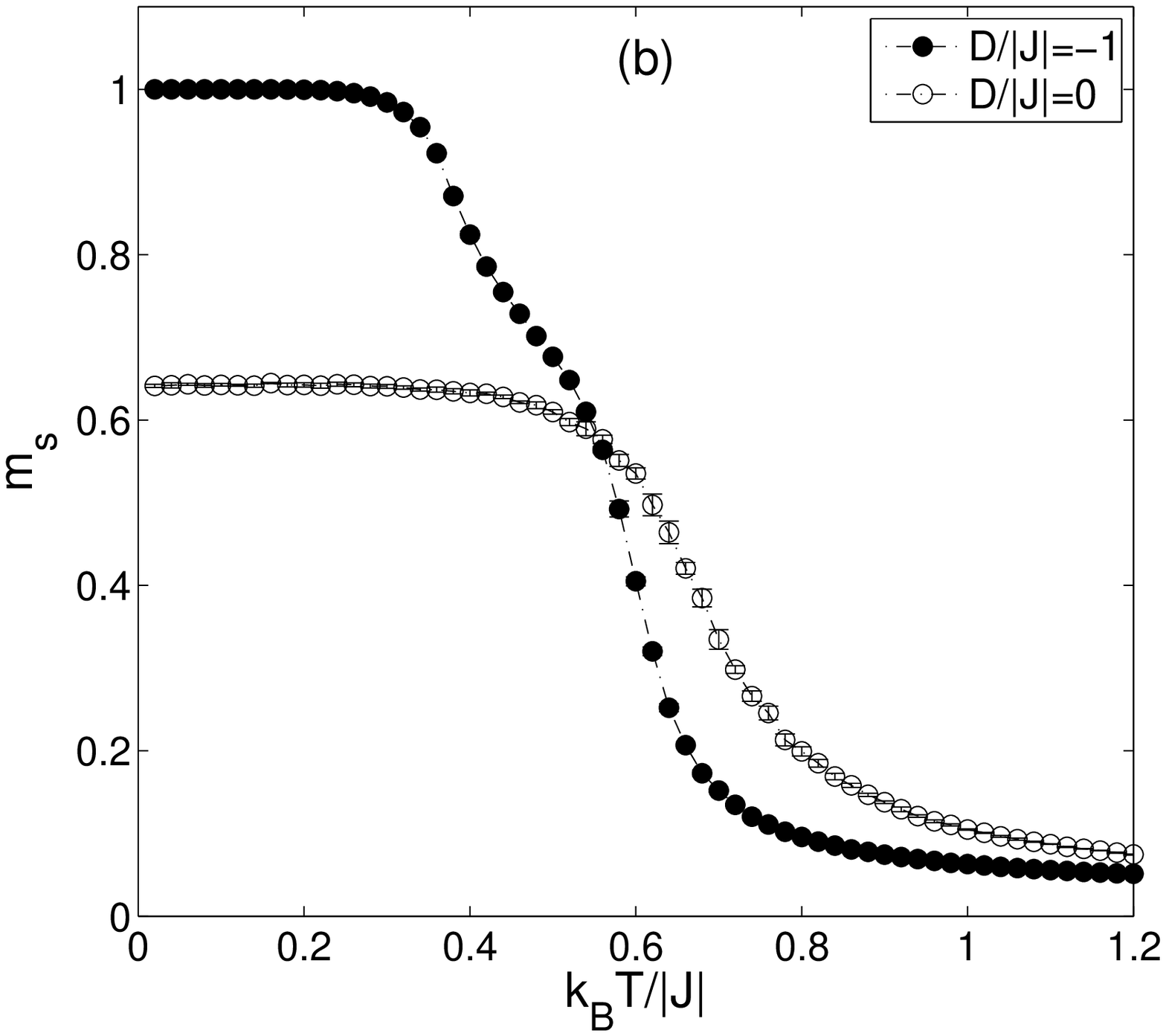}\label{fig:ms-T_L48}}
    \subfigure{\includegraphics[scale=0.42]{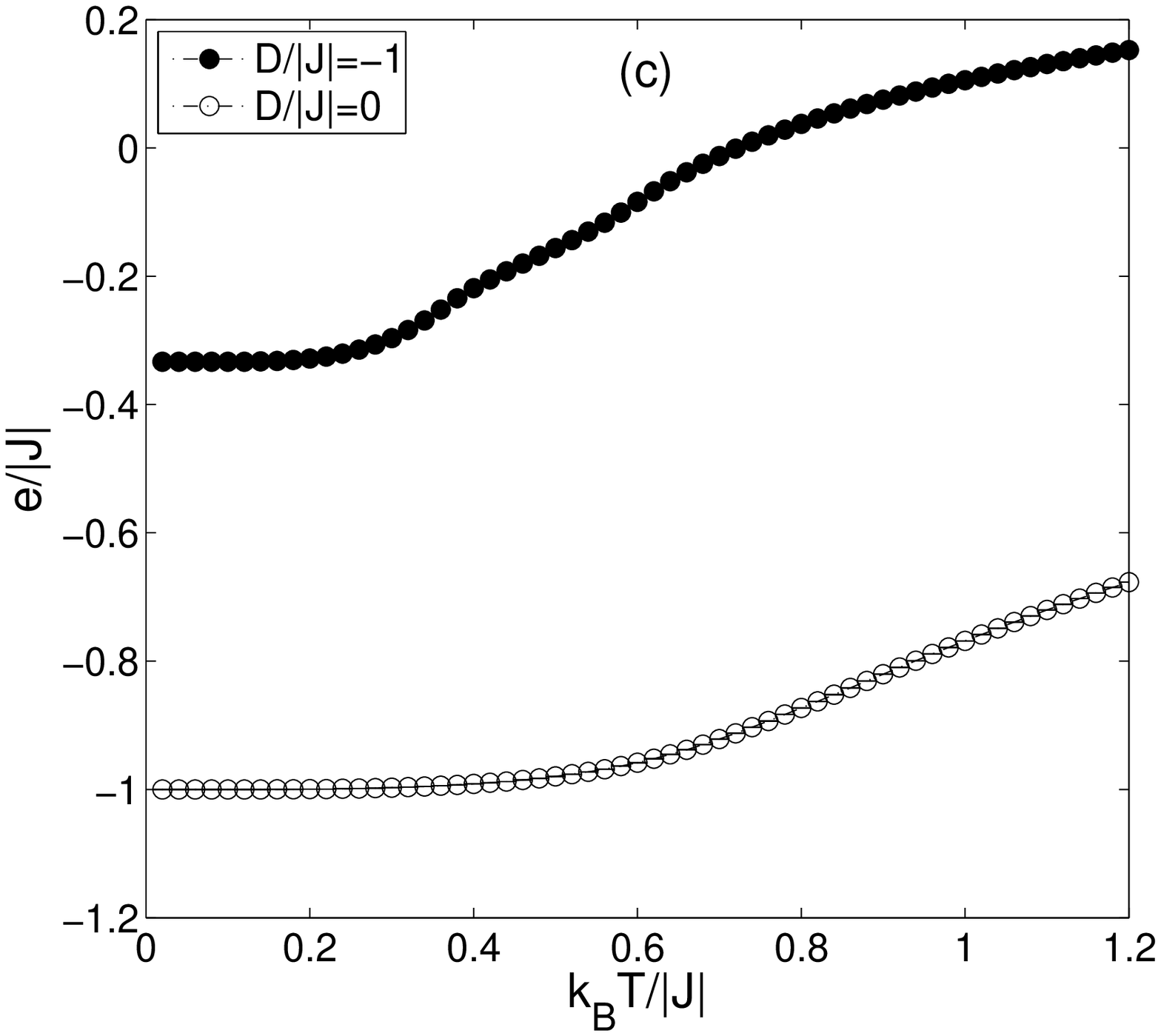}\label{fig:e-T_L48}}\\
    \subfigure{\includegraphics[scale=0.42]{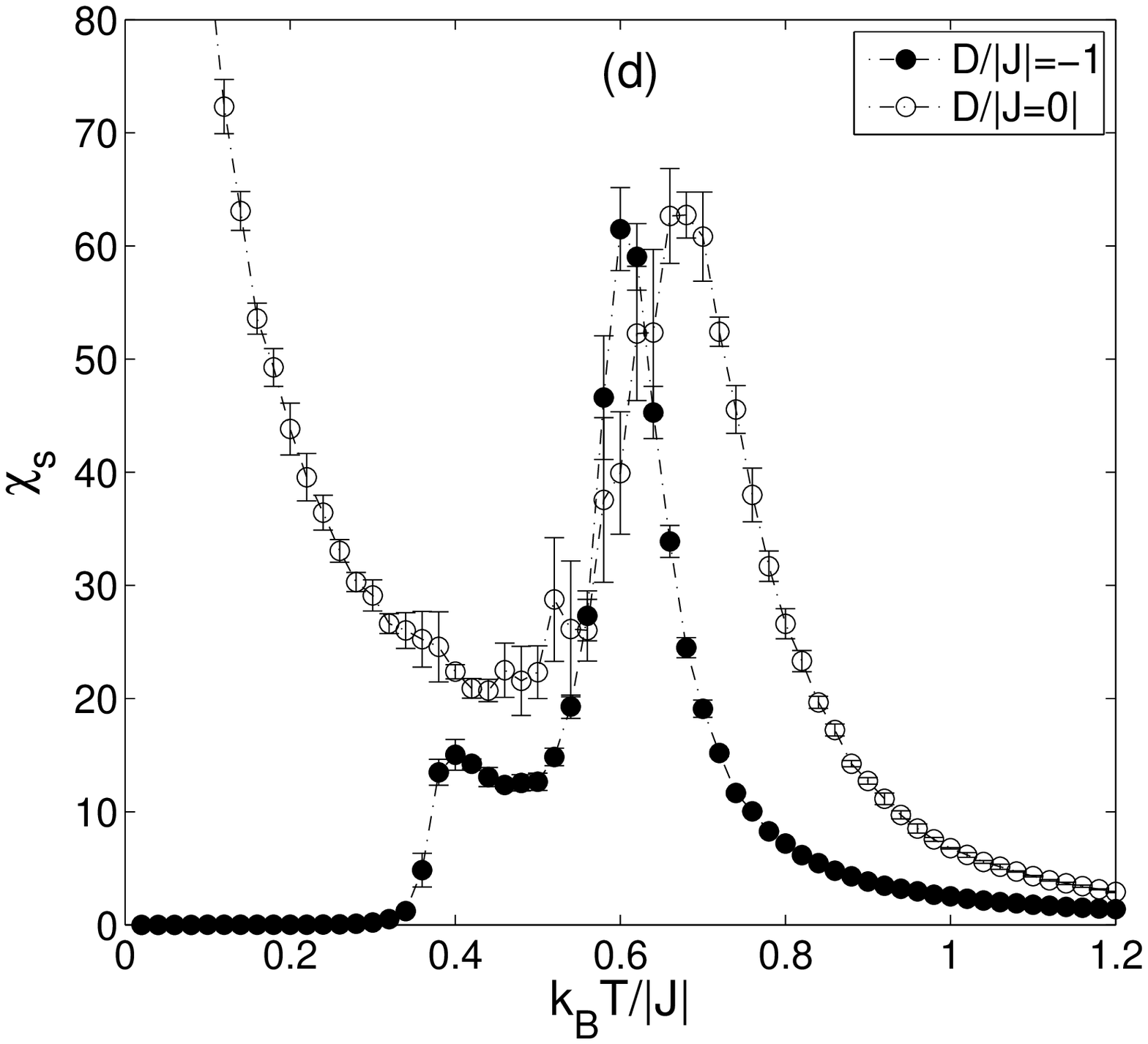}\label{fig:xis-T_L48}}
    \subfigure{\includegraphics[scale=0.42]{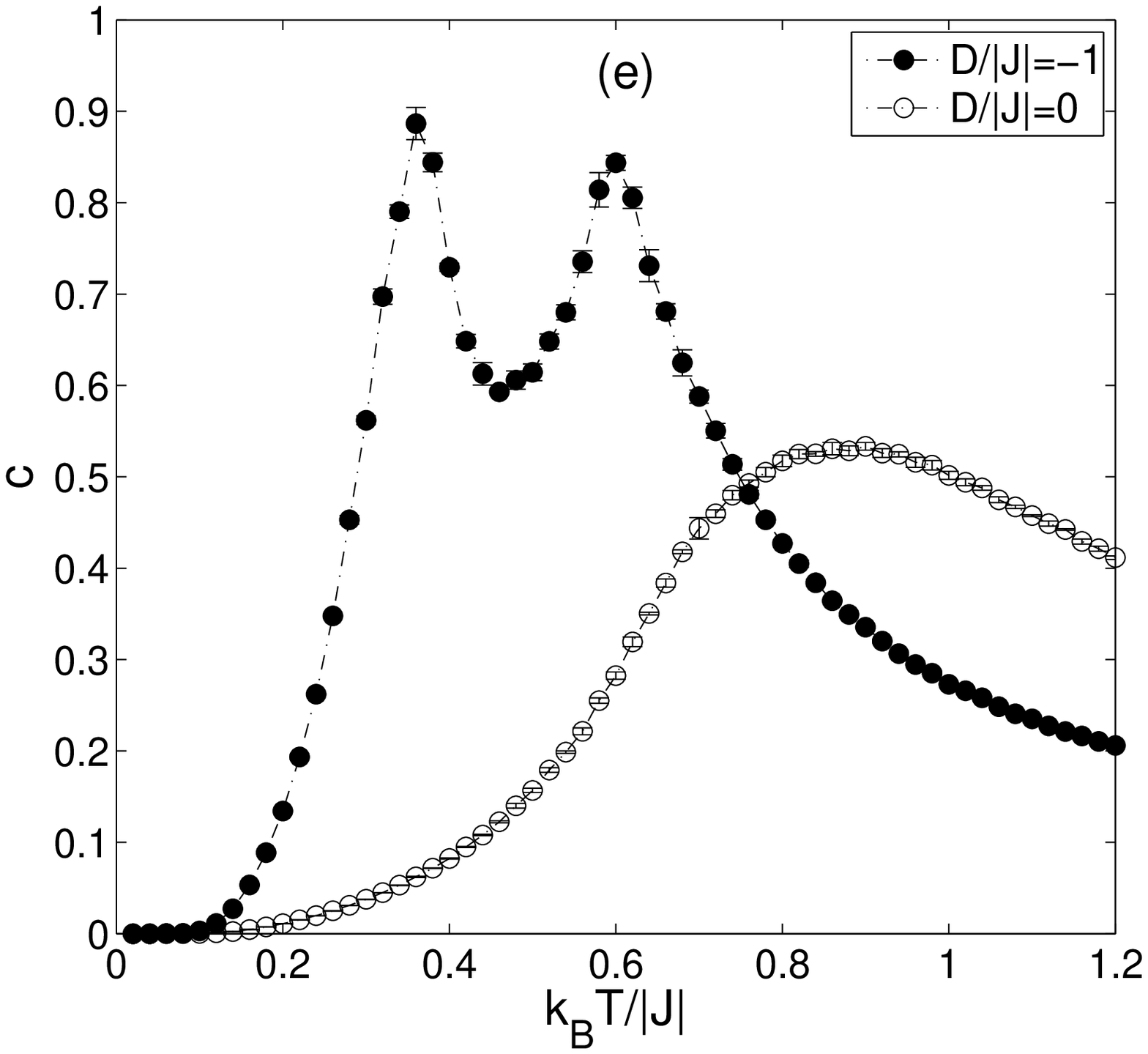}\label{fig:c-T_L48}}
\caption{Temperature variation of (a) the sublattice magnetizations $m_{\mathrm{A}}$, $m_{\mathrm{B}}$, and $m_{\mathrm{C}}$, (b) the staggered magnetization $m_s$, (c) the internal energy $e/|J|$, (d) the staggered susceptibility $\chi_s$ and (e) the specific heat $c$, for $D/|J|=-1$ and $0$.}
\label{fig:xxx-T_L48}
\end{figure}

In order to confront the behavior of the system with no single-ion anisotropy, which is expected to show only the BKT phase transition~\cite{yama}, with the long-range order behavior predicted to appear within the range $3J/2<D<0$, in Fig.~\ref{fig:xxx-T_L48} we plot the temperature dependencies of some relevant thermodynamic quantities for two values of $D/|J|=0$ and $-1$, representing the two cases. The behavior for $D/|J|=0$ is clearly different from that observed for $D/|J|=-1$. While the former shows only one anomaly at higher temperatures, the latter displays an additional anomaly at lower temperatures. In particular, as observed in the behaviors of the sublattice magnetizations and the staggered magnetization (order parameter), presented in Figs.~\ref{fig:mi-T_L48} and~\ref{fig:ms-T_L48}, respectively, in both cases some degree of ordering is initiated in two sublattices at higher temperatures. However, unlike the case with no anisotropy, for $D/|J|=-1$ this initial increase in the quantities $m_{\mathrm{A}}$, $m_{\mathrm{B}}$ and $m_s$ is followed by another increase to the fully saturated values at lower temperatures. As the temperature is lowered, one and two anomalies are also observed in the behavior of the internal energy, tending to the expected ground-state values of $e/|J|=-1$ and $-1/3$, for the cases of $D/|J|=0$ and $D/|J|=-1$, respectively (Fig.~\ref{fig:e-T_L48}). The anomalies in the staggered magnetization and the internal energy are reflected in the corresponding number of peaks in the staggered susceptibility and the specific heat, shown in Figs.~\ref{fig:xis-T_L48} and ~\ref{fig:c-T_L48}, respectively.   

\begin{figure}[t]
\centering
    \includegraphics[scale=0.45]{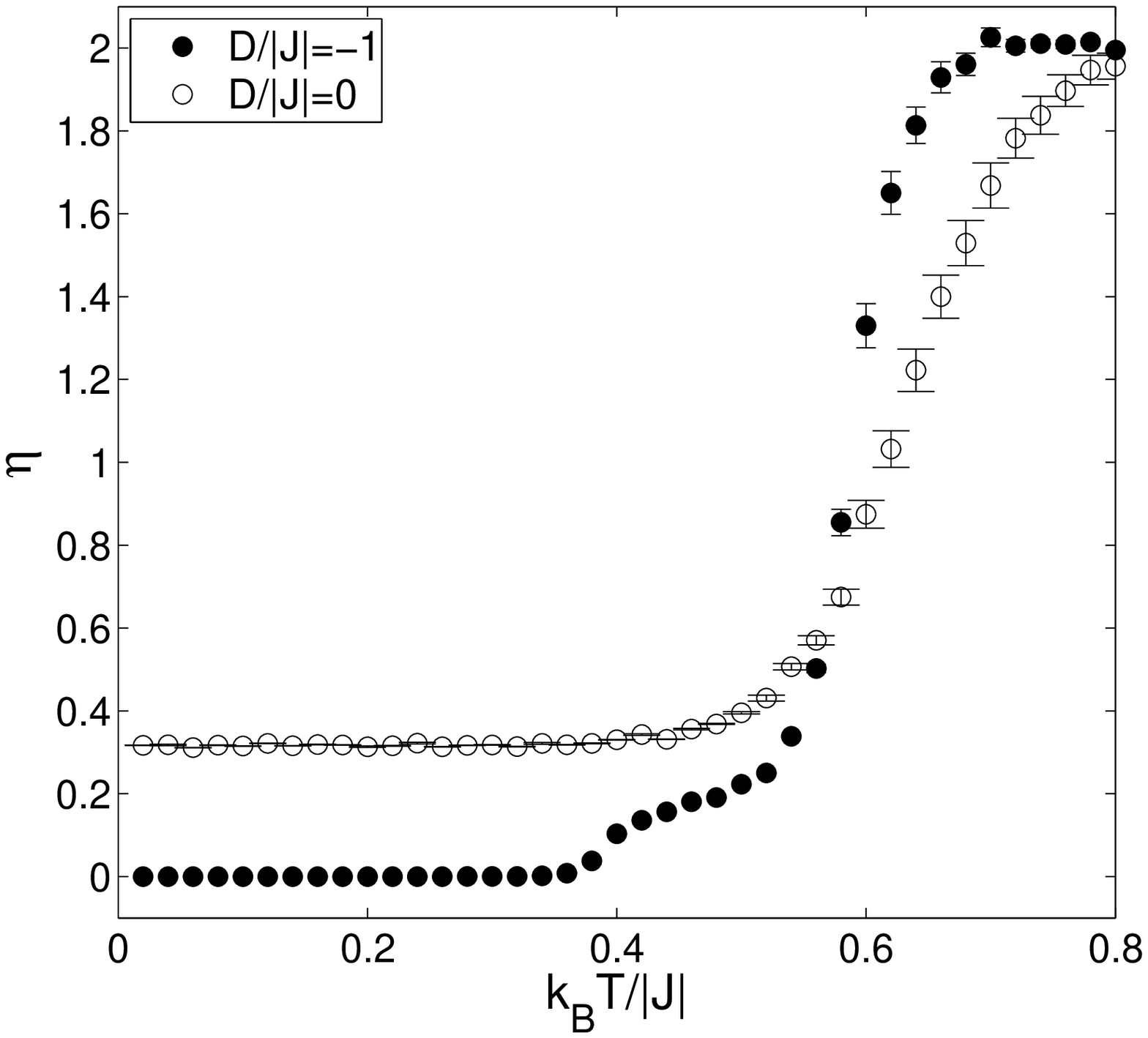}
\caption{Temperature variation of the exponent $\eta$, for $D/|J|=-1$ and $0$. The symbols and the error bars respectively represent the mean and the extreme values obtained from Eqs.~(\ref{ms_FSS}) and~(\ref{Y_FSS}).}
\label{fig:eta-T_L48_D0}
\end{figure}

The above anomalies in the thermodynamic quantities suggest occurrence of transitions between different phases. To identify the respective phases, we employ FSS analyses of the order parameter $m_s$ and the quantity $Y$, defined by the relations~(\ref{ms_FSS}) and~(\ref{Y_FSS}), respectively. Temperature variation of the scaling exponent $\eta$ is presented in Fig.~\ref{fig:eta-T_L48_D0}. The value of $\eta=0$, observed for $D/|J|=-1$, means that the system displays at sufficiently low temperatures (below $k_BT/|J| \approx 0.35$) long-range order~\cite{naga,yama}. On the other hand, for $D/|J|=0$, the value of $\eta$ remains finite and in the low temperature region levels off at the value of $\eta \approx 0.317$, in agreement with the previous ground-state investigations~\cite{naga,lipo}.

\begin{figure}[t]
\centering
    \subfigure[$D/|J|=1$]{\includegraphics[clip,scale=0.28]{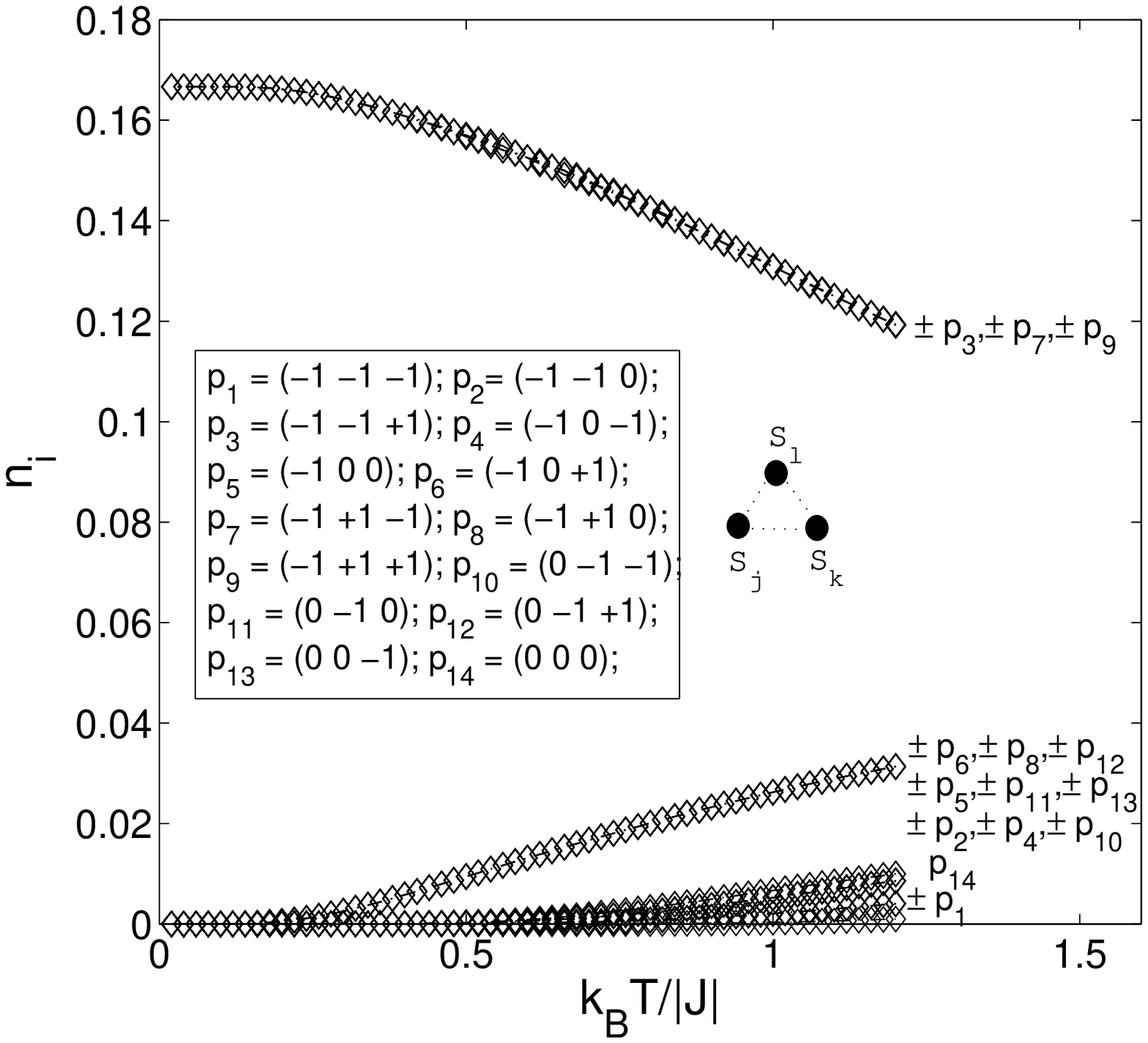}\label{fig:dos-T_L48_D+1}}
    \subfigure[$D/|J|=0$]{\includegraphics[clip,scale=0.28]{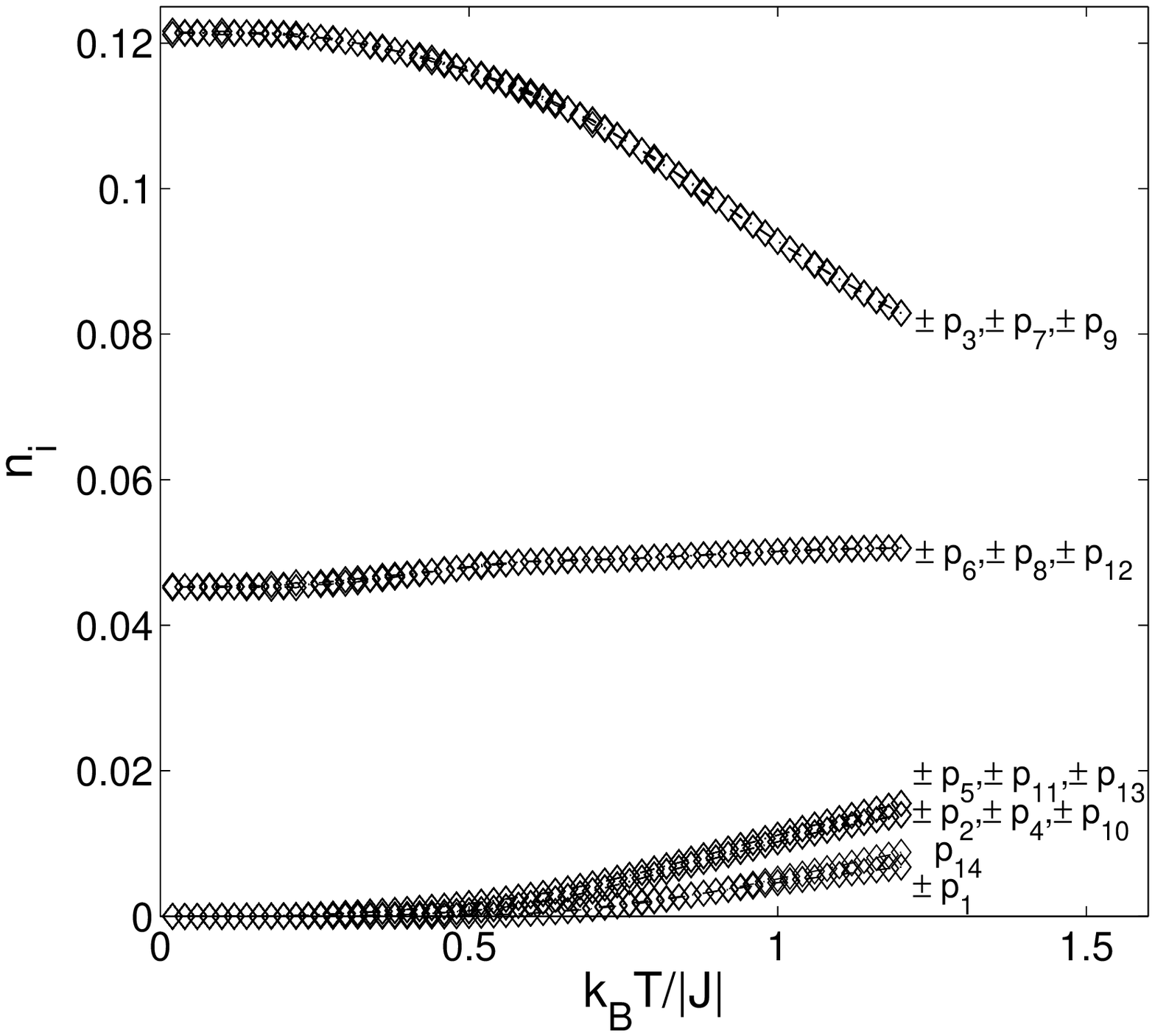}\label{fig:dos-T_L48_D0}}
    \subfigure[$D/|J|=-1$]{\includegraphics[clip,scale=0.28]{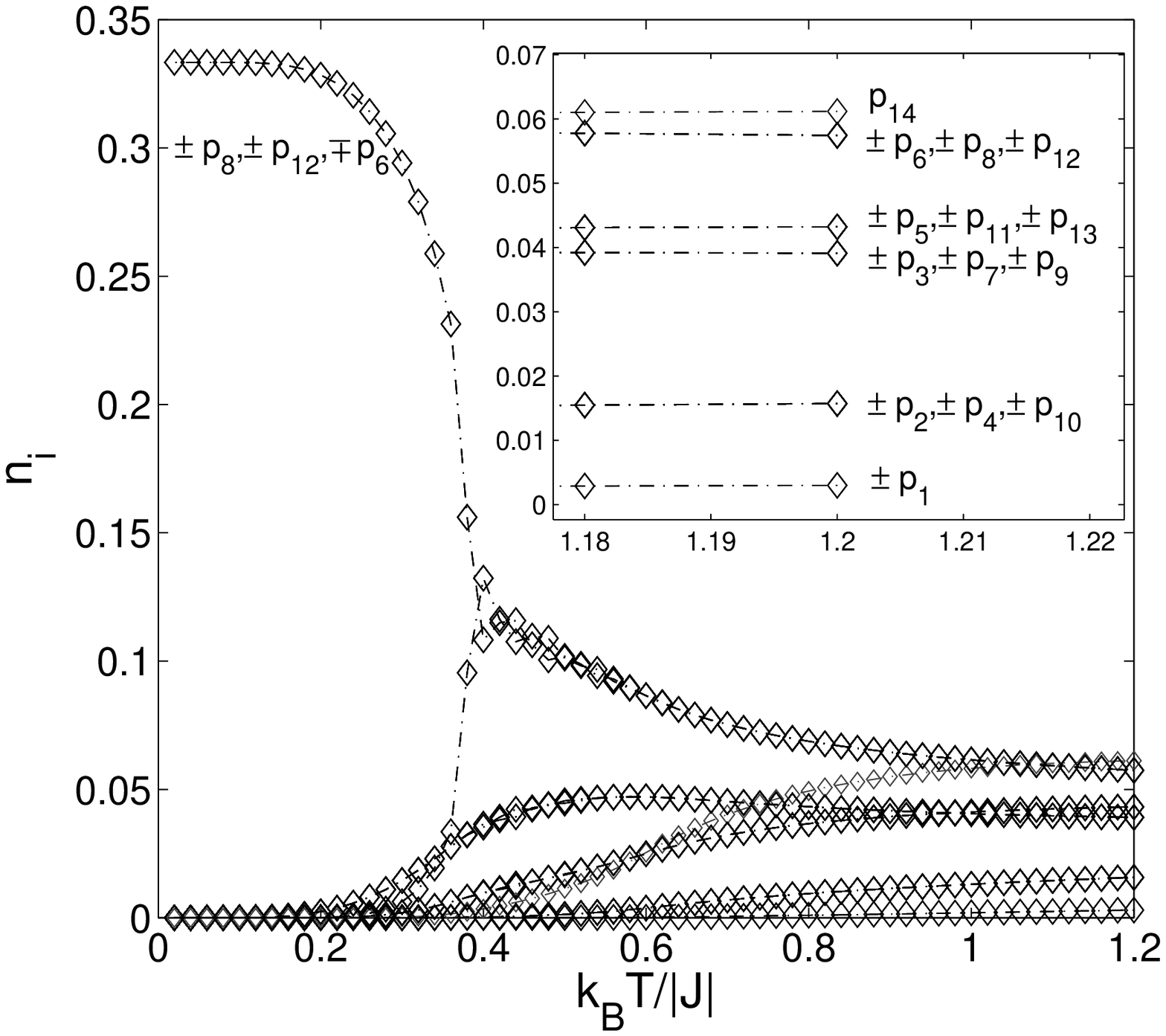}\label{fig:dos-T_L48_D-1}}\\
    \subfigure[$D/|J|=1$]{\includegraphics[clip,scale=0.33]{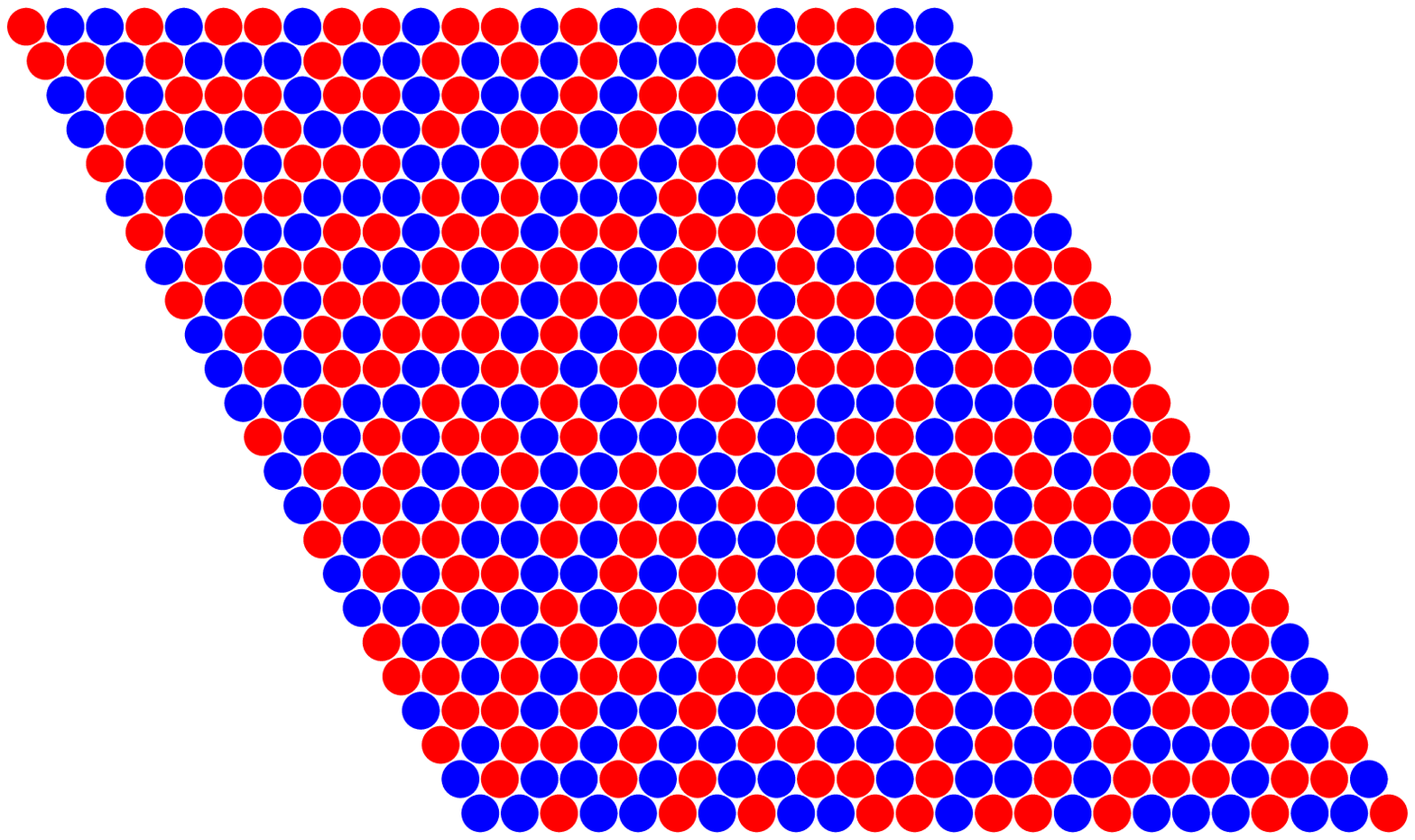}\label{fig:snap_T0_D+1_L24}}
    \subfigure[$D/|J|=0$]{\includegraphics[clip,scale=0.33]{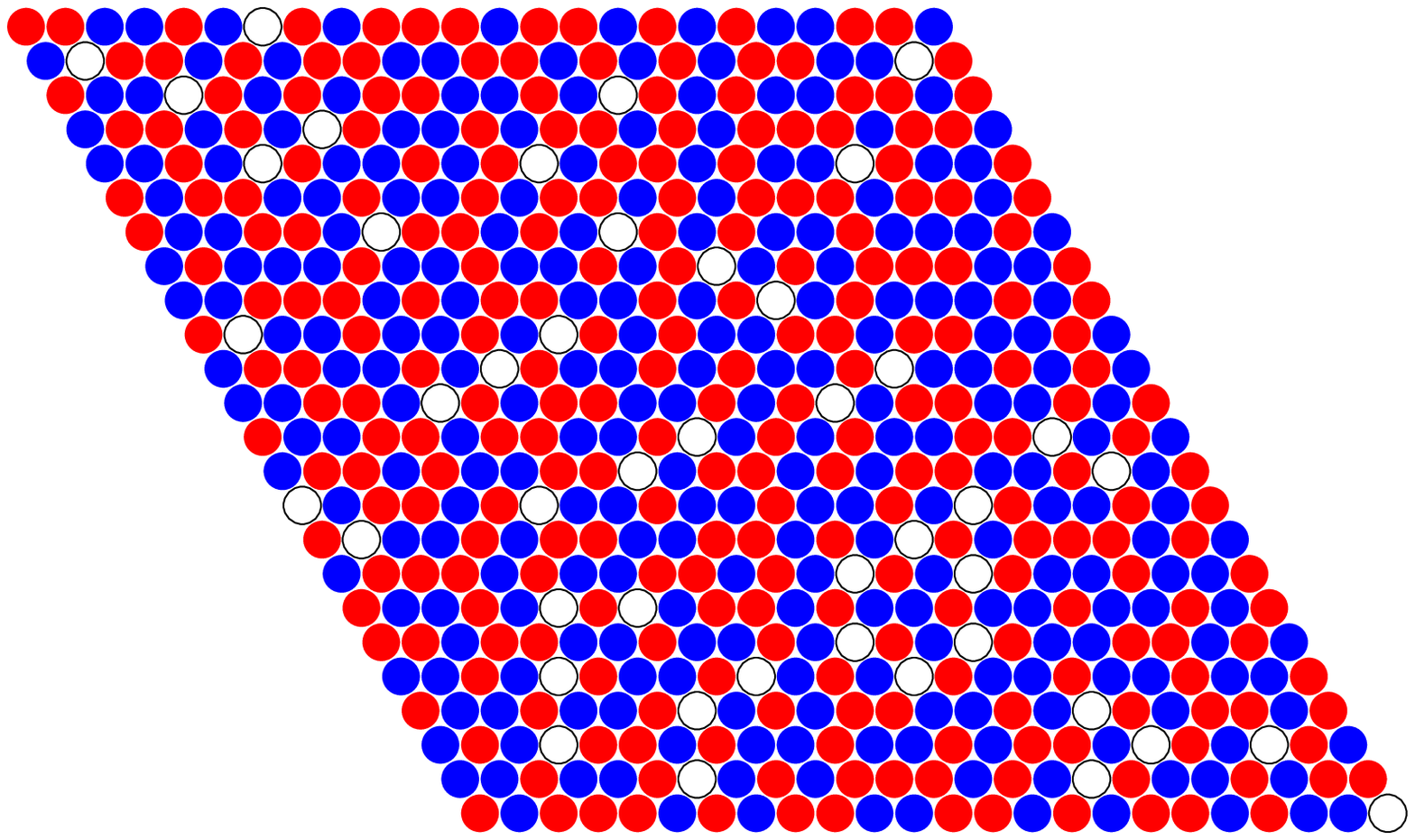}\label{fig:snap_T0_D0_L24}}
    \subfigure[$D/|J|=-1$]{\includegraphics[clip,scale=0.33]{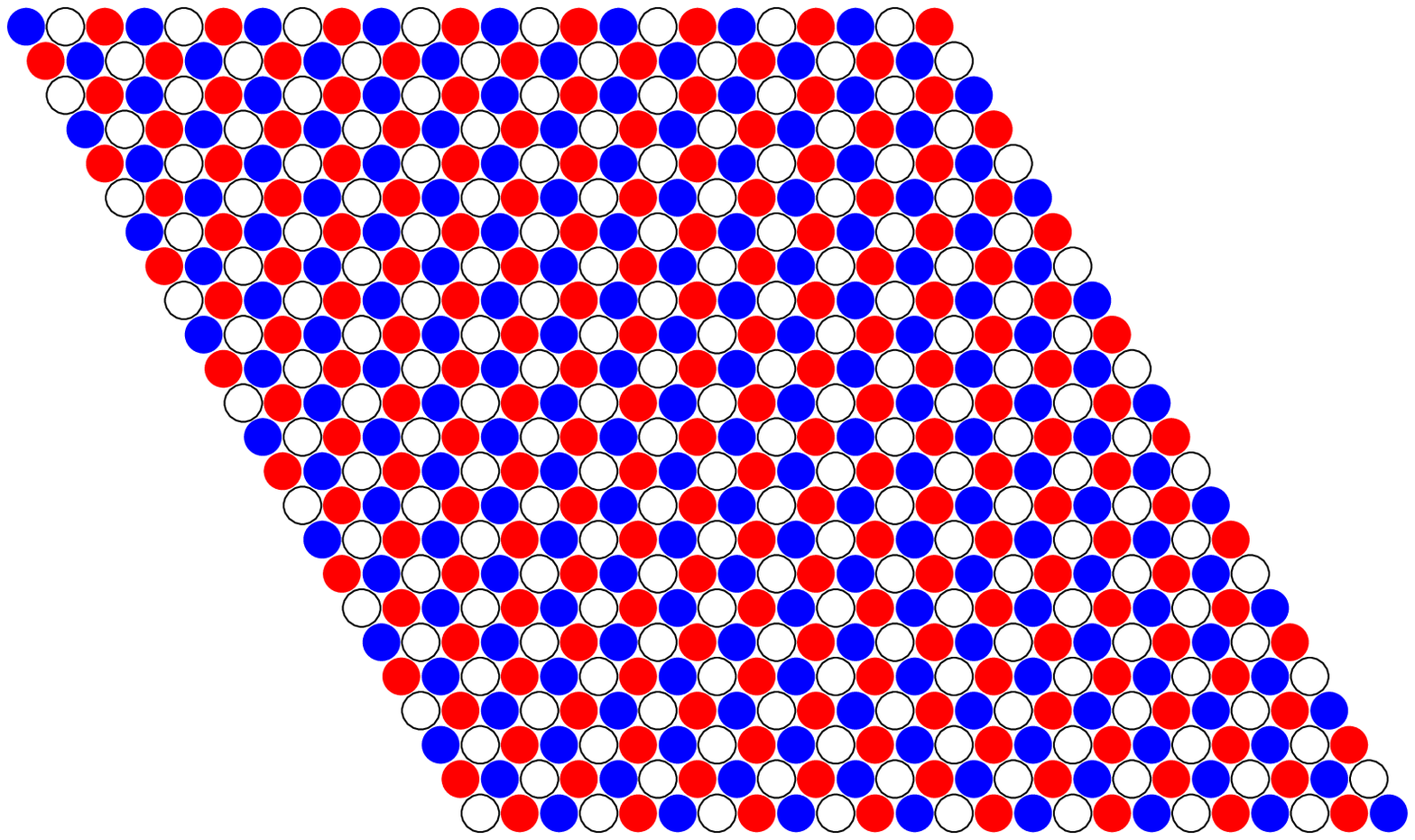}\label{fig:snap_T0_D-1_L24}}
\caption{(Color online) Temperature variation of the densities of local patterns $p_i=(S_j,S_k,S_l)$, $i=1-14$, (upper row) and the corresponding ground-state spin configuration snapshots (lower row), for $D/|J|=1,0$ and $-1$. The red (light gray), blue (dark gray) and white circles represent the spin states $-1$, $+1$ and $0$, respectively.}
\label{fig:dos-T_L48}
\end{figure}

Finally, we found it interesting to look into the system's behavior also at microscopic level. More specifically, we studied the snapshots of spin configurations at different temperatures both visually and quantitatively. For the latter we introduced the quantity, which we termed density of local patterns $n_i$, which is defined as a relative number of different spin patterns $p_i=(S_j,S_k,S_l)$, $i=1-14$~\footnote{In fact, there are in total 27 patterns, however, the remaining 13 patterns can be obtained by the spin flipping, i.e., as $-p_i$.}, on the elementary triangular plaquettes observed in the simulated states. The temperature variations of $n_i$ for different values of the single-ion anisotropy strength are shown in Figs.~\ref{fig:dos-T_L48_D+1}-\ref{fig:dos-T_L48_D-1}, with all the patterns $p_i$ listed in Fig.~\ref{fig:dos-T_L48_D+1}. The snapshots of the respective spin configurations in the ground state are also presented in Figs.~\ref{fig:snap_T0_D+1_L24}-\ref{fig:snap_T0_D-1_L24}~\footnote{For clarity we show the snapshots for a smaller lattice size of $L=24$.}. For comparison we also included the case with positive value of $D/|J|=1$. For $D/|J|=1$, the degenerate spin patterns which sum to $\pm 1$ and include no zeros, i.e., $\{\pm p_3,\pm p_{7},\pm p_{9}\}$, are equally represented and, in accordance with the above ground-state arguments, for $t \rightarrow 0$ are the only ones present in the system. On the other hand, for $D/|J|=-1$, due to broken symmetry the system in the partially ordered phase chooses one the two configurations involving either the set of patterns $\{p_6,-p_{8},-p_{12}\}$ or $\{-p_6,p_{8},p_{12}\}$. Again, in agreement with the ground-state arguments, for $t \rightarrow 0$ the patterns of either set are equally represented and there are no other patterns in the system. In between the two phases for $D/|J|=0$, the patterns $\{\pm p_3,\pm p_{7},\pm p_{9}\}$ and $\{\pm p_6,\mp p_{8},\mp p_{12}\}$ mix together since they contribute with the same energy. However, as evidenced from both Figs.~\ref{fig:dos-T_L48_D0} and~\ref{fig:snap_T0_D0_L24}, they do not appear with the same probabilities. Namely, the patterns $\{\pm p_3,\pm p_{7},\pm p_{9}\}$ systematically prevail in all the simulations started from various initial states and in the ground state they are almost $2.7$-times more represented than the patterns $\{\pm p_6,\mp p_{8},\mp p_{12}\}$. This can be understood by looking at larger patterns, namely, hexagonal ones formed by a central spin and its six nearest neighbors. Then, only in the patterns with the so called ``free'' central spins (i.e., the spins with the nearest neighbors of alternating signs) the state of the central spins is irrelevant with the respect to the energy and, therefore, the states $+1$, $-1$ and $0$ are equally probable. Therefore, the number of the triangular patterns $\{\pm p_6,\mp p_{8},\mp p_{12}\}$ is proportional to the number of the hexagonal patterns with the ``free'' central spins in the configurations such as the one shown in Fig.~\ref{fig:snap_T0_D+1_L24}, noting that the zero-state central spin generates six triangles with the patterns $\{\pm p_6,\mp p_{8},\mp p_{12}\}$.      

\section{CONCLUSIONS}
We studied the geometrically frustrated spin-1 triangular Ising antiferromagnet in the presence of a single-ion anisotropy. We showed that within the range of negative values of the anisotropy parameter $D/|J| \in (-1.5,0)$ the system displays a partial long-range order in the low-temperature region and another phase with algebraically decaying correlation function at higher temperatures. For $D/|J| < -1.5$ the system is in a non-magnetic state and for $D/|J| > 0$ it behaves like a spin-1/2 model, with no long-range ordering due to frustration. The ground-state phase transition at $D/|J| = 0$ is characterized by the coexistence of the energetically equivalent patterns of the types $(\pm 1 \pm 1 \mp 1)$ and $(0 \pm 1 \mp 1)$ at a constant ratio of approximately $27:10$.

\begin{acknowledgments}
This work was supported by the Scientific Grant Agency of Ministry of Education of Slovak Republic (Grant VEGA No. 1/0234/12). The authors acknowledge the financial support by the ERDF EU (European Union European regional development fund) grant provided under the contract No. ITMS26220120005 (activity 3.2.).
\end{acknowledgments}

\end{document}